\documentclass[aps,prl,twocolumn,showpacs,superscriptaddress,floatfix,nofootinbib,10pt]{revtex4-1}

\usepackage{graphicx}
\usepackage{amsmath}
\usepackage{epsfig}
\usepackage{helvet}
\usepackage{amssymb}

\newcommand{\be}{\begin{equation}}
\newcommand{\ee}{\end{equation}}
\newcommand{\bea}{\begin{eqnarray}}
\newcommand{\eea}{\end{eqnarray}}

\newcommand{\tr}{\mbox{tr}}
\newcommand{\bra}[1]{\mbox{$\langle #1 |$}}
\newcommand{\ket}[1]{\mbox{$| #1 \rangle$}}

\def\tr{ \mbox{tr}}

\begin{document}

\title{A class of highly entangled many-body states that can be efficiently simulated}

\author{G. Evenbly}
\affiliation{Institute for Quantum Information, California Institute of Technology, MC 305-16, Pasadena CA 91125, USA}
\email{evenbly@caltech.edu}
\author{G. Vidal}
\affiliation{Perimeter Institute for Theoretical Physics, Waterloo, Ontario N2L 2Y5, Canada}  \email{gvidal@perimeterinstitute.ca}
\date{\today}

\begin{abstract}
We describe a quantum circuit that produces a highly entangled state of $N$ qubits from which one can efficiently compute expectation values of local observables. This construction yields a variational ansatz for quantum many-body states that can be regarded as a generalization of the multi-scale entanglement renormalization ansatz (MERA), and to which we refer as the branching MERA. In a lattice system in $D$ dimensions, the scaling of entanglement of a region of size $L^{D}$ in the branching MERA is not subject to restrictions such as a boundary law $L^{D-1}$, but can be proportional to the size of the region, as we demonstrate numerically for $D=1,2$ dimensions.
\end{abstract}

\maketitle

Simulating quantum many-body systems with a classical computer is generically hard, in that the computational resources needed to describe the state of $N$ quantum systems (e.g. $N$ quantum spins) grow exponentially with $N$. However, in some appropriate limit, quantum systems effectively become the classical entities of our everyday world. There are several good reasons to investigate the conditions under which quantum many-body systems can be efficiently simulated by classical means. At a fundamental level, the cost of classically simulating a many-body system may be used to define an operational notion of quantumness or classicality. In addition, such studies shed light on where the power of quantum computation \cite{QC} lies, while they may guide the development of new quantum algorithms. At a more practical level, identifying classes of many-body states that can be efficiently simulated classically has led to a host of new tools \cite{DMRG,TEBDforMB,PEPS,ER} to investigate quantum many-body systems, which are of interest in a wide range of research areas including condensed matter physics, high energy physics, quantum chemistry, nanotechnology, etc.

Examples of quantum evolutions that can be efficiently simulated include those that entangle only a constant number $n$ of the $N$ systems at a time \cite{Jozsa}; and those that generate a restricted amount of entanglement between any left-right bipartition of the $N$ systems when ordered in one dimension \cite{TEBDforQC}. Another important set of simulable quantum evolutions is obtained by generating it with a restricted, non-universal set of transformations (e.g. Clifford group \cite{Clifford}, and free boson and fermion evolutions, generated by quadratic operators \cite{free}). Finally, a quantum circuit on $N$ two-level systems or qubits, can also be simulated classically, in the sense that expectation values of local operators can be efficiently computed, by restricting the causal structure of the circuit, as is the case of the multi-scale entanglement renormalization ansatz (MERA) \cite{MERA}, which is used as a variational ansatz for the ground states of many-body Hamiltonians. In the quantum circuit for the MERA, constraints on the causal structure result in a restricted amount of entanglement in the many-body state. For instance, in a lattice in $D\geq 2$ dimensions, the entanglement entropy $S_{L^{D}}$ of a region of $L^{D}$ sites scales as $L^{D-1}$, and thus obeys a boundary law. However, the ground states of certain many-body systems display violations of the area law, such as the scaling $L^{D-1}\log D$ \cite{logFermions,logBosons}, and can therefore not be represented by the MERA.

\begin{figure}
  \begin{centering}
\includegraphics[width=7.5cm]{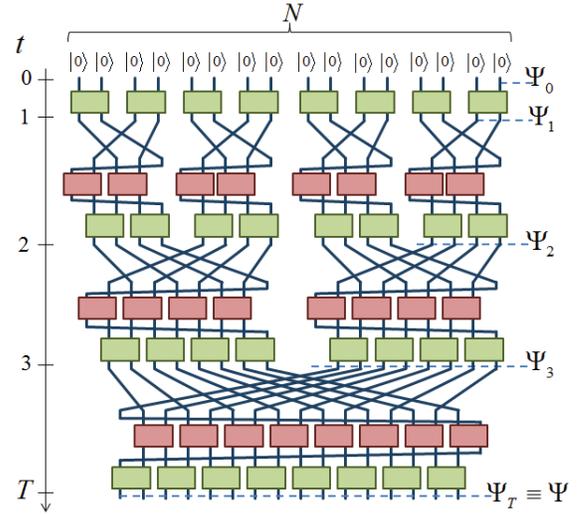}  \end{centering}
  \caption{(Color online) Quantum circuit that transforms the product state $\ket{0}^{\otimes N}$ into an entangled state $\ket{\Psi}$ of $N$ qubits. It contains $O(N\log (N))$ generic gates (and $O(N^2)$ swap gates) organized in about $T=\log_2(N)$ horizontal layers, and $O(N)$ vertical branches. The circuit is self-similar and has a peculiar causal structure, such that the expectation value $\bra{\Psi}\hat{O}\ket{\Psi}$ of a local operator $\hat{O}$ can be efficiently computed.}
  \label{fig:branchingMERA}
\end{figure}

Here we introduce a quantum circuit where, as in the MERA, the causal structure is restricted in such a way that one can efficiently compute expectation values for local operators. However, in contrast with the MERA, the scaling of entanglement entropy can reproduce violations of the boundary law, all the way up to a bulk law, $L^D$, which we demonstrate numerically for $D=1,2$ dimensions. We start by describing the circuit and proving its simulability in $D=1$ dimensions, with the $D>1$ extension being straightforward.

\textit{Quantum Circuit.---} We consider a family of quantum circuits on $N=2^{T}$ (for a positive integer $T$) qubits or quantum wires, and $O(N\log(N))$ generic unitary gates acting on two nearest-neighbor wires, supplemented with $O(N^2)$ swap gates that exchange the position of two wires. Figure \ref{fig:branchingMERA} shows the quantum circuit for $T=4$. A discrete time $t$, $t \in \{0, 1, 2, \cdots, T \equiv \log_2 (N)\}$, organizes the generic two-qubit gates into (double) layers. The circuit transforms the initial unentangled state $\ket{0}^{\otimes N}$ of $N$ qubits at time $t=0$ into an $N$-qubit entangled state $\ket{\Psi}$ at time $t = T$,
\begin{equation}\label{eq:circuit}
    \ket{0}^{\otimes N} \rightarrow \ket{\Psi}.
\end{equation}

\begin{figure}
  \begin{centering}
\includegraphics[width=8.0cm]{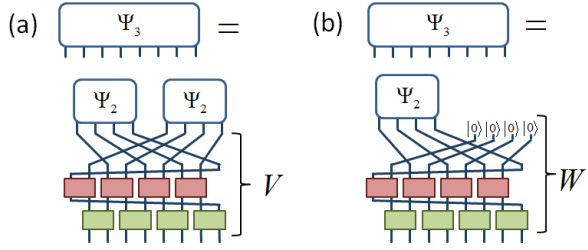}  \end{centering}
  \caption{(Color online) (a) Computation of state $\ket{\Psi_{t+1}}$ from two copies of state $\ket{\Psi_t}$ (for $t=2$), by means of a unitary transformation $V$ made of swaps and two-qubit gates. (b) Computation of state $\ket{\Psi_{t+1}}$ from state $\ket{\Psi_t}$ (for $t=2$), by means of an isometric transformation $W$.}
  \label{fig:VW}
\end{figure}

\begin{figure}
  \begin{centering}
\includegraphics[width=7.0cm]{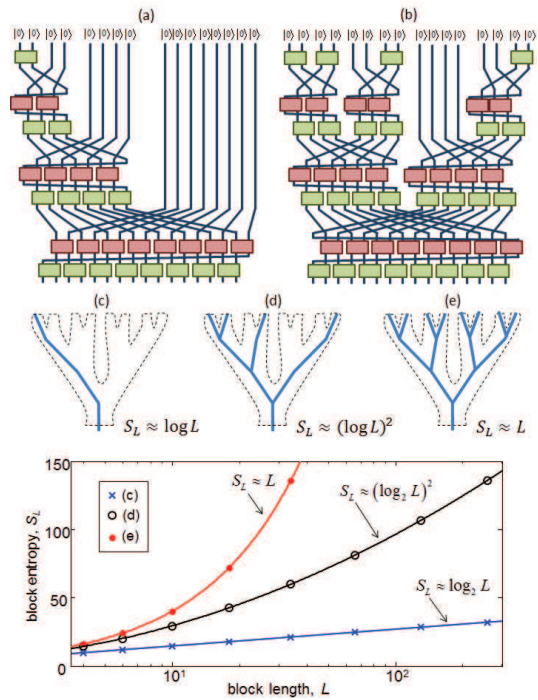}  \end{centering}
  \caption{(Color online) (a) Quantum circuit corresponding to the MERA. (b) Quantum circuit corresponding to a branching MERA with less branches than that it Fig. \ref{fig:branchingMERA}. (c)-(e) Holographic trees for the circuits in Figs. \ref{fig:branchingMERA2}(a)-(b) and Fig. \ref{fig:branchingMERA}. The plot shows the scaling of entanglement entropy in a block of $L$ sites evaluated numerically from a branching MERA created with random, free fermionic gates. Solid lines display a best fit over the indicated functional form.}
  \label{fig:branchingMERA2}
\end{figure}

 The quantum circuit has a branching, self-similar structure such that, for instance, the circuit for $N$ qubits is obtained by merging two circuits for $N/2$ qubits, and so on. For notational simplicity, first we will describe the action of the circuit assuming that the gates on any two of its branches are the same (in Lemma 2 below, we will lift this assumption). At $t=0$, the $N$ wires are grouped into $N/2$ nearest neighbor pairs and, for each pair, the product state $\ket{\Psi_0}\otimes \ket{\Psi_0}$ (where $\ket{\Psi_0} \equiv \ket{0}$) is transformed into an entangled state $\ket{\Psi_1}$ of the two wires, so that applying the first row of gates has the net effect of transforming $N$ copies of $\ket{\Psi_0}$ into $N/2$ copies of $\ket{\Psi_{1}}$, $\ket{\Psi_0}^{\otimes N} \rightarrow \ket{\Psi_1}^{\otimes N/2}$. In general, at time $t$, a number of swaps and a (double) layer of two-qubit gates act on pairs of sets of $2^{t}$ wires, where each set of wires is in an entangled state $\ket{\Psi_{t}}$, to produce an entangled state $\ket{\Psi_{t+1}}$ of $2^{t+1}$ wires,
\begin{equation}\label{eq:double}
\ket{\Psi_{t}} \otimes \ket{\Psi_{t}} \rightarrow \ket{\Psi_{t+1}} = V \left( \ket{\Psi_{t}} \otimes \ket{\Psi_{t}} \right),
\end{equation}
see Fig. \ref{fig:VW}(a), so that from time $t$ to time $t+1$ the state of the $N$ wires evolves as
\begin{equation}\label{eq:theta}
    \ket{\Psi_{t}}^{\otimes N/2^{t}} \rightarrow \ket{\Psi_{t+1}}^{\otimes N/2^{t+1}}.
\end{equation}
Thus, the circuit produces the sequence of states
\begin{equation}\label{eq:sequence}
    \ket{\Psi_0}^{\otimes N} \rightarrow \ket{\Psi_1}^{\otimes N/2} \rightarrow \cdots \rightarrow \ket{\Psi_{T-1}}^{\otimes 2} \rightarrow \ket{\Psi_{T}},
\end{equation}
where $\ket{\Psi_{T}} \equiv \ket{\Psi}$ is the state of the $N$ outgoing wires of the quantum circuit. For each value of $t$, state $\ket{\Psi_t}$ is a highly entangled state of $2^t$ qubits, see Fig. \ref{fig:branchingMERA2}.

\begin{figure}
  \begin{centering}
\includegraphics[width=8.0cm]{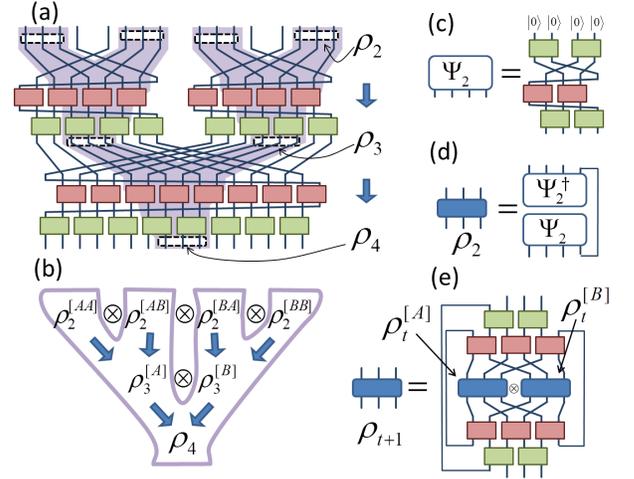}  \end{centering}
  \caption{(Color online) (a) The past causal cone of three outgoing wires, indicated by the shaded region of the circuit, has the shape of a binary tree with $2^{s}$ branches at time $t=T-s$. In each of these branches, at any given time $t$ it only involves three contiguous wires. (b) Accordingly, the reduced density matrix $\rho_T\equiv \rho$ (here, $T=4$), can be obtained by computing a sequence of $T-1$ reduced density matrices on three wires (here, $\{\rho_2, \rho_3, \rho_4\}$), see Eq. \ref{eq:sequenceRho}. (c)-(d) $\rho_2$ in terms of a small tensor network. (e) Computation of $\rho_{t+1}$ from $\rho_t^{\otimes 2}$ by applying gates and tracing out three wires.}
  \label{fig:causalCone}
\end{figure}

\textit{Efficient computation of local expectation values.---}
A key property of the quantum circuit above is that the expectation value $\bra{\Psi}\hat{O}\ket{\Psi}$ of any local operator $\hat{O}$ (which for convenience we assume to be supported on three contiguous outgoing wires of the circuit\cite{moreThanThree}) can be computed efficiently. This is in sharp contrast with the same computation for a generic state of $N$ qubits, which would have a cost exponential in $N$. This remarkable result follows from the two lemmas below and the equality $\bra{\Psi}\hat{O}\ket{\Psi} = \tr (\rho \hat{O})$, where $\rho$ is the reduced density matrix for three contiguous outgoing wires of the quantum circuit with $N(=2^{T})$ wires, obtained by tracing out the other $N-3$ outgoing wires. To discuss computational costs with more generality, we consider that each wire corresponds to a $\chi$-dimensional quantum system, with $\chi=2$ for qubits.

\textit{Lemma 1 (particular case).---} When any two branches of the circuit have the same gates, the reduced density matrix $\rho$ can be obtained with computational time and space scaling as $O(\chi^{12} \log (N))$ and $O(\chi^{8})$, respectively.

\textit{Proof.---} Let us consider the past causal cone of three contiguous out-going wires, defined as the subset of wires and gates that can influence the reduced density matrix $\rho$, see Fig \ref{fig:causalCone}(a). If we cut this causal cone into time slices, then at $t=T$ we find the three out-going wires; at time $t=T-1$ there are six wires; in general, at time $t=T-s$ there are (at most) $3\times 2^s$ wires. We will obtain $\rho$ by computing the state of these sets of wires, progressing from $t=0$ all the way to $t=T$. Importantly, the factorization of Eq. \ref{eq:sequence} for the state of all the wires implies that the density matrices inside the causal cone (for $t\geq 2$) factorize analogously as
\begin{eqnarray}\label{eq:sequenceRho}
    \rho_2^{\otimes N/4} \rightarrow  \rho_3^{\otimes N/8} \rightarrow \cdots
    \cdots \rightarrow \rho_{T-1}^{\otimes 2} \rightarrow \rho_{T},
\end{eqnarray}
where $\rho_t$ is a reduced density matrix of three contiguous wires at time $t$, and $\rho_{T} \equiv \rho$, see Fig. \ref{fig:causalCone} (b). Therefore $\rho$ can be simply obtained by computing the sequence of $T-1$ reduced density matrices $\{\rho_{2}, \rho_{3}, \cdots, \rho_{T}\}$. The first density matrix $\rho_2$ is obtained from the entangled state $\ket{\Psi_{2}}$ of four wires by tracing out one of them, as explained in Figs. \ref{fig:causalCone}(c)-(d). This can be seen to require time $O(\chi^6)$ and memory $O(\chi^4)$. Then, for $t=2, \cdots, T-1$, the density matrix $\rho_{t+1}$ is obtained from two copies of $\rho_{t}$, as indicated in Fig. \ref{fig:causalCone}(e). This requires computational time and memory that scale as $O(\chi^{12})$ and $O(\chi^8)$, respectively. Thus, the overall computational time and memory grow with $\chi$ and $N$ as $O(\chi^{12} \log (N))$ and $O(\chi^8)$ respectively, completing the proof.

\textit{Lemma 2 (general case).---} When all the gates in the circuit are allowed to be different, the reduced density matrix $\rho$ can be obtained with computational time and space scaling as $O(\chi^{12} N)$ and $O(\chi^{8}\log(N))$ \cite{Transform}.

\textit{Proof.--} The proof is analogous to that of Lemma 1. At time $t$ the state of the $N$ wires still factorizes as $N/2^{t}$ states of $2^t$ wires as in Eq. \ref{eq:sequence}, but each set of $2^t$ wires is now in a different state. Consequently, at time $t$ the reduced density matrix inside the past causal cone still factorizes as $N/2^{t}$ density matrices of three wires as in Eq. \ref{eq:sequenceRho}, but now each set of three wires is in a different mixed state. One such density matrix at time $t+1$ is obtained by combining a pair of density matrices at time $t$, as in Fig. \ref{fig:causalCone}(b) (where branches are labeled by a sequence of $A$'s and $B$'s), and this has to be done for $N/2^{t+1}$ pairs. Adding the computational time for all values of $t$, we see that $\rho$ can be computed with time and extra memory scaling as $O(\chi^{12}N)$ and $O(\chi^8 \log(N))$.

\textit{Variational many-body ansatz.---} Consider now a one-dimensional lattice $\cal{L}$ made of $N$ sites, where each site corresponds to a qubit or, more generally, a $\chi$-level system, and a local Hamiltonian $\hat{H} = \sum_l \hat{H}_l$ that decomposes as a sum of local terms $\hat{H}_l$ acting on at most a block of three contiguous sites including site $l$. A natural application of the quantum circuit of Fig. \ref{fig:branchingMERA} is to use the resulting state $\ket{\Psi}$ to define a variational class of states for the $N$-site lattice $\mathcal{L}$, which we could use, e.g. to approximate the ground state of Hamiltonian $\hat{H}$. We refer to this ansatz as the branching MERA, for reasons that will become clear shortly. The variational parameters are contained in the gates of the quantum circuit (excluding the swaps), with each gate depending on $O(\chi^4)$ continuous parameters. In the general case of lemma 2 there there are $O(N\log(N))$ independent gates [down to $O(N)$ in the particular case of lemma 1]. Therefore, $\ket{\Psi}$ depends on $O(\chi^4 N \log (N))$ [respectively, $O(\chi^4 N)$] variational parameters \cite{scaleInvariant}. For any choice of these parameters, we can efficiently compute the energy, $E_{\Psi} \equiv \bra{\Psi} \hat{H} \ket{\Psi} = \sum_l \bra{\Psi} \hat{H}_l \ket{\Psi}$, since we can efficiently compute each of the $O(N)$ terms $\bra{\Psi} \hat{H}_l \ket{\Psi}$. The variational parameters in the branching MERA can then be optimized, e.g. by adapting the optimization scheme of Ref. \cite{MERAalgorithms}.

The regular MERA, see Fig. \ref{fig:branchingMERA2} (a), can be regarded as a particular case of the branching MERA, in which the gates for all but one branch are set to be the identity gate. Notice that in the regular MERA, at time $t$ an entangled state $\ket{\Psi_t}$ of $2^{t}$ wires and the product state $\ket{0}^{\otimes 2^{t}}$ of $2^{t}$ fresh wires are mapped into an entangled state $\ket{\Psi_{t+1}}$ of $2^{t+1}$ wires,
\begin{equation}\label{eq:double2}
\ket{\Psi_{t}} \rightarrow \ket{\Psi_{t+1}} = W \ket{\Psi_{t}},
\end{equation}
where the isometric transformation $W$ is obtained from the unitary transformation $V$ by fixing the state of $2^{t}$ incoming wires to $\ket{0}$, $W \equiv V \ket{0}^{\otimes 2^{t}}$, see Fig. \ref{fig:VW}(b). In this case, a local expectation value $\bra{\Psi}\hat{O}\ket{\Psi}$ is also obtained by computing a sequence of reduced density matrices contained in the past causal cone of the support of the local operator $\hat{O}$, but now the causal cone contains only one branch, and therefore Eq. \ref{eq:sequenceRho} is replaced with the simplified sequence
\begin{eqnarray}\label{eq:sequenceRhoMod}
    \rho_2 \rightarrow  \rho_3 \rightarrow \cdots
    \cdots \rightarrow \rho_{T-1} \rightarrow \rho_{T}.
\end{eqnarray}

\begin{figure}
  \begin{centering}
\includegraphics[width=7.5cm]{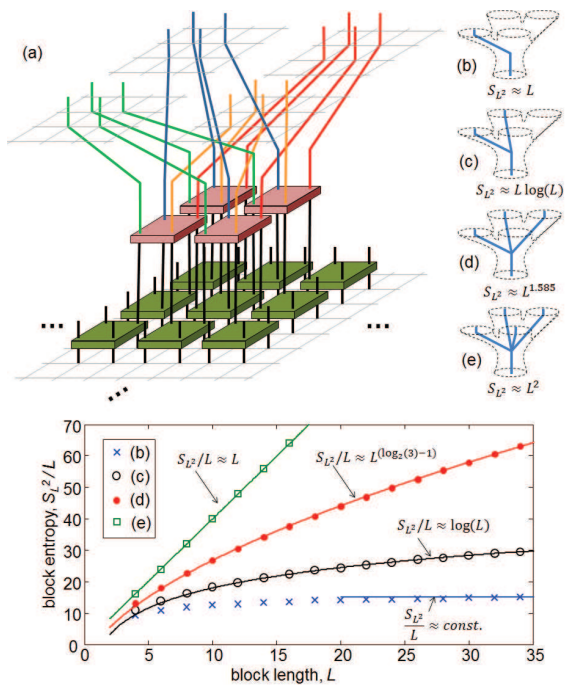}  \end{centering}
  \caption{(Color online) (a) Local detail of the branching MERA in $D=2$ dimensions. Four different subcircuits are combined into one by means of a double layer of four-wire gates. (b)-(e) As in $D=1$ dimensions, one may remove some of the branches by setting some of the incoming wires to a fixed state $\ket{0}$. This results in holographic trees with a branching factor ranging from one (regular MERA, (b)) to four (maximal branching, (e)). The plot shows the scaling of entanglement entropy in a block of $L^2$ sites evaluated numerically from a branching MERA created with random, free fermionic gates. Solid lines display a best fit over the indicated functional form.}
  \label{fig:2D}
\end{figure}

From a condensed matter perspective, one step of the regular MERA (from time $t+1$ to time $t$) can be regarded as implementing a coarse-graining (isometric) transformation $W^{\dagger}$ that maps the state $\ket{\Psi_{t+1}}$ of a lattice of $2^{t+1}$ sites into the state $\ket{\Psi_{t}}$ of a coarse-grained lattice of $2^{t}$ sites, and thus implements a real-space renormalization group transformation \cite{ER}.
In contrast, one (reversed) step of the branching MERA can be regarded as implementing a decoupling (unitary) transformation $V^{\dagger}$ that maps the state $\ket{\Psi_{t+1}}$ of a lattice of $2^{t+1}$ sites into the decoupled state $\ket{\Psi_{t}}\otimes \ket{\Psi_{t}}$ of two coarse-grained lattices of $2^{t}$ sites each, thus disentangling two sets degrees of freedom that become decoupled at a given length/energy scale \cite{decoupling}.

More generally, setting a subset of branches to the identity, as in the example of Fig. \ref{fig:branchingMERA2}(b), produces other interesting subclasses of many-body states, all of which can be efficiently manipulated. These different subclasses of branching MERA can be roughly characterized by their holographic tree \cite{decoupling}, which represents their branching structure, see Fig. \ref{fig:branchingMERA2}(c)-(e). For a generic choice of variational parameters, the holographic tree largely characterizes the structure of entanglement in the branching MERA. For instance, focusing on the entanglement entropy $S_{L}$ of a block of $L$ contiguous sites ($L \leq N/2$), we find that different holographic trees imply different scaling of $S_{L}$, see Fig. \ref{fig:branchingMERA2}. This scaling ranges from a boundary law, where $S_{L}$ saturates to a constant, all the way up to a bulk law of the branching MERA of Fig. \ref{fig:branchingMERA}, where $S_L$ grows as $\approx L$; and it includes, as particular examples, the logarithmic scaling $S_L \approx  \log (L)$ characteristic of the regular MERA in Fig. \ref{fig:branchingMERA2}(a), as well as a less orthodox scaling $S_L \approx \log^2(L)$ of the branching MERA in Fig. \ref{fig:branchingMERA2}(b). 

The generalization of the branching MERA to $D\geq 2$ space dimensions is straightforward, see Fig. \ref{fig:2D}, and results in an ansatz where the scaling of the entanglement entropy of a block of $L^{D}$ sites also ranges, depending on the holographic tree, anywhere between a boundary law $S_{L^D} \approx L^{D-1}$ and a bulk law $S_{L^D} \approx L^D$, including logarithmic violations of the boundary law, $S_{L^D} \approx L^{D-1} \log L$. The latter corresponds to the scaling of entanglement entropy in certain compressible gapless phases in $D\geq 2$ dimensions, including Fermi liquids \cite{logFermions} and spin-Bose metals \cite{logBosons}, suggesting that the branching MERA may offer a suitable representation for those phases.


\begin{thebibliography}{99}
\bibitem{moreThanThree} The reduced density matrix for $n$ contiguous wires, as well as $n$-point correlators, can also be computed with a cost that scales exponentially in $n$ but is still efficient in the number $N$ of wires.
\bibitem{Transform} Lemma 2 also applies to the state $\ket{\Psi [i_1 i_2 \cdots i_N]}$ obtained by replacing the initial state $\ket{0}^{\otimes N}$ of the $N$ qubits in Fig. \ref{fig:branchingMERA} with $\ket{i_1 i_2 \cdots i_N} \equiv \ket{i_1}\otimes \ket{i_2}\otimes \cdots \otimes \ket{i_N}$, where $[i_1, i_2, \cdots, i_N]$ is a string of zeroes and ones. Therefore, the circuit transforms the (unentangled) computational basis $\{ \ket{i_1 i_2 \cdots i_N} \}$ of the Hilbert space of $N$ qubits, into a highly entangled basis $\{ \ket{\Psi [i_1 i_2 \cdots i_N]} \}$. This transformation involves all length scales, and may be thought of as some form of generalized Fourier transform in the space of $N$ qubits.
\bibitem{scaleInvariant} Another interesting case is when all the gates are repetition of a finite set of gates, in which case $O(\chi^4)$ parameters completely specifies the ansatz. Notice that then we can consider the thermodynamic limit, $N \rightarrow \infty$. An efficient computation of local expectation values is still possible by considering the fixed-point(s) of the non-linear map $\rho_t^{\otimes 2} \rightarrow \rho_{t+1}$ in Fig. \ref{fig:causalCone}(e).
\end{thebibliography}
\end{document}